\documentclass[preprint,endfloats*,superscriptaddress]{revtex4-1}

\usepackage{graphicx}
\usepackage{dcolumn}
\usepackage{bm}
\usepackage{amsmath}
\usepackage{latexsym}

\newcommand{\rfig}[1]{Fig.~\ref{#1}}
\newcommand{\rfigs}[1]{Figs.~\ref{#1}}

\newcommand{\rref}[1]{Ref.~[\onlinecite{#1}]}
\newcommand{\req}[1]{Eq.~(\ref{#1})}
\newcommand{\reqs}[1]{Eqs.~(\ref{#1})}

\makeindex

\begin{document}

\title{Magnetotransport induced by anomalous Hall effect}

\author{Jiaji Zhao}
\author{Bingyan Jiang}
\affiliation{State Key Laboratory for Artificial Microstructure and Mesoscopic Physics, Frontiers Science Center for Nano-optoelectronics, Peking University, Beijing 100871, China}
\author{Jinying Yang}
\affiliation{Beijing National Laboratory for Condensed Matter Physics, Institute of Physics, Chinese Academy of Sciences, Beijing 100190, China}
\affiliation{School of Physical Sciences, University of Chinese Academy of Sciences, Beijing 100049, China}
\author{Lujunyu Wang}
\author{Hengjie Shi}
\author{Guang Tian}
\affiliation{State Key Laboratory for Artificial Microstructure and Mesoscopic Physics, Frontiers Science Center for Nano-optoelectronics, Peking University, Beijing 100871, China}
\author{Zhilin Li}
\author{Enke Liu}
\affiliation{Beijing National Laboratory for Condensed Matter Physics, Institute of Physics, Chinese Academy of Sciences, Beijing 100190, China}
\author{Xiaosong Wu}
\email{xswu@pku.edu.cn}
\affiliation{State Key Laboratory for Artificial Microstructure and Mesoscopic Physics, Frontiers Science Center for Nano-optoelectronics, Peking University, Beijing 100871, China}
\affiliation{Collaborative Innovation Center of Quantum Matter, Beijing 100871, China}
\affiliation{Shenzhen Institute for Quantum Science and Engineering, Southern University of Science and Technology, Shenzhen 518055, China}
\affiliation{Peking University Yangtze Delta Institute of Optoelectronics, Nantong, Jiangsu 226010, China}

\begin{abstract}
In a magnetic metal, the Hall resistance is generally taken to be the sum of the ordinary Hall resistance and the anomalous Hall resistance. Here it is shown that this empirical relation is no longer valid when either the ordinary Hall angle or the anomalous Hall angle is not small. Using the proper conductivity relation, we reveal an unexpected magnetoresistance (MR) induced by the anomalous Hall effect (AHE). A $B$-linear MR arises and the sign of the slope depends on the sign of the anomalous Hall angle, giving rise to a characteristic bowtie shape. The Hall resistance in a single-band system can exhibit a nonlinearity which is usually considered as a characteristic of a two-band system. A $B$-symmetric component appears in the Hall. These effects reflect the fundamental difference between the ordinary Hall effect and the AHE. Furthermore, we experimentally reproduce the unusual MR and Hall reported before in Co$_3$Sn$_2$S$_2$ and show that these observations can be well explained by the proposed mechanism. MR often observed in quantum anomalous Hall insulators provides further confirmation of the picture. The effect may also account for the large MR observed in non-magnetic three-dimensional topological Dirac semimetals.
\end{abstract}



\maketitle

\section{Introduction}

Although the anomalous Hall effect was discovered over a century ago, there has been a long-time controversy over its mechanism \cite{Nagaosa2010}. Significant progress was made upon the introduction of the Berry phase \cite{Xiao2010}. It is now well understood that the intrinsic anomalous Hall effect (AHE) is determined by the integral of the Berry curvature over occupied states \cite{Nagaosa2010}. With the advent of the topological band concept, the AHE has attracted revived interest. The strong Berry curvature appearing in topological bands can give rise to a large AHE \cite{Chang2013,Suzuki2016Dec,Kuroda2017Nov,Ghimire2018,Kim2018Sep,Belopolski2019Sep,Noky2020,Deng2020Feb}. Currently, great efforts have been made in a quest for a strong AHE \cite{Suzuki2016Dec,Shekhar2018,Liu2018,Wang2018,Li2020Jul,Fujishiro2021Jan,Jiang2021Mar}. A related thermoelectric effect, the anomalous Nernst effect (ANE), has also been actively pursued \cite{Guin2019a,Sakai2018,Sakai2020,Sakuraba2020Apr}. Achieving a large AHE and ANE will potentially lead to efficient spintronic and energy conversion devices. The strength of AHE may be characterized by the anomalous Hall angle, defined as $\vartheta=\arctan(\sigma_{xy}^\mathrm{A}/\sigma_{xx})$, where $\sigma_{xy}^\mathrm{A}$ is the anomalous Hall conductivity, and $\sigma_{xx}$ is the longitudinal conductivity. In the past, $\tan\vartheta$ was less than 0.1 \cite{Kim2018Sep,Liu2018,Li2020Jul}. It was quickly boosted to 0.33 in topological materials \cite{Shen2020Aug}. It is likely that we will witness a sharp increase of $\tan\vartheta$ in the near future.

The Hall resistivity measured in experiments is influenced by both ordinary Hall effect (OHE) and AHE. An empirical relation is widely used to extract the AHE resistivity, that is,
\begin{equation}
\rho_{yx}=R_\mathrm{H}^\mathrm{O} B_z + \rho_{yx}^\mathrm{A}, 
\label{eq.empirical}
\end{equation}
where $R_\mathrm{H}^\mathrm{O}$ is the ordinary Hall effect coefficient and $B_z$ is the perpendicular field. The second term represents the anomalous Hall resistivity, which is proportional to the perpendicular magnetization. This relation applies to many materials \cite{Nagaosa2010}. It is shown in this paper that \req{eq.empirical} is valid only when the ordinary Hall angle and the anomalous Hall angle are small, which is well satisfied in conventional materials. However, it is no longer the case in light of recent progress in finding a large anomalous Hall angle. In a magnetic Weyl semimetal, Co$_3$Sn$_2$S$_2$, $\tan\vartheta=0.33$ has been achieved \cite{Shen2020Aug}. In the meantime, unusual magnetoresistance (MR) and Hall effects were observed and mechanisms were proposed to explain these effects \cite{Yang2020b,Zeng2021Jul}. We reproduce these experimental observations and show that they can be well explained by AHE, without invoking any additional mechanism. Our results emphasize the necessity of employing the correct conductivity relation for analyzing AHE data when the Hall angle is not small. Moreover, they suggest an alternative explanation for the large MR observed in topological Dirac semimetals \cite{Liang2015Mar,Novak2015Jan,Feng2015Aug,Tang2019May}.

\section{Experiments}

Two types of Co$_3$Sn$_2$S$_2$ single crystal were measured in this work. Both were grown by a chemical vapor transfer method. One is bulk crystals with a typical size of $1100 \times 500 \times 60\ \mathrm{\mu m}^3$ \cite{Jiang2021Jun}. The other is a nanoflake with a thickness of 86~nm \cite{Yang2020b}. Bulk crystals were cut and polished. Silver paste was used for electrical contacts. For the nanoflake sample, standard electron-beam lithography and argon plasma etching were employed for the fabrication of Hall bars. Au films with Ti as an adhesion layer were used in metallization for electrical contacts. A standard low-frequency alternating-current method was employed for electrical transport measurements using a lock-in amplifier.

\section{Results and discussion}

Let us consider the simplest case, that is, an isotropic single band. In the absence of an AHE, the resistivity is expressed as a second rank tensor
\begin{equation}
\rho=\rho_0
\begin{pmatrix}
  1 & -\tan\theta\\
  \tan\theta & 1
\end{pmatrix},
\label{eq.rho}
\end{equation}
where $\rho_0$ and $\theta$ are the zero-field Drude resistivity and the ordinary Hall angle, respectively. $\tan\theta=\mu B$, where $\mu$ is the carrier mobility. To include the AHE, it is worth noting that the AHE gives rise to a contribution to the conductivity rather than the resistivity. For instance, the intrinsic anomalous Hall conductivity is determined by the integral of the Berry curvature over occupied states and is thus independent of the longitudinal resistivity \cite{Nagaosa2010}. Therefore, the total conductivity is the sum of the conductivity given by \req{eq.rho} and the anomalous Hall conductivity $\sigma_{xy}^\mathrm{A}$
\begin{equation}
\sigma=
\begin{pmatrix}
  \sigma_0/(1+\tan^2\theta) & \sigma_0\tan\theta/(1+\tan^2\theta)+\sigma_{xy}^\mathrm{A}\\
  -\sigma_0\tan\theta/(1+\tan^2\theta)-\sigma_{xy}^\mathrm{A} & \sigma_0/(1+\tan^2\theta)
\end{pmatrix},
\label{eq.sigma}
\end{equation}
where $\sigma_0=1/\rho_0$. Define $\sigma_{xy}^\mathrm{A}\equiv \rho_{yx}^\mathrm{A}/\rho_0^2$ and let $\theta=0$ (zero field). It can be seen that $\tan\vartheta=\sigma_{xy}^\mathrm{A}/\sigma_0=\rho_{yx}^\mathrm{A}/\rho_0$, as expected. Plugging it into \req{eq.sigma}, one finally obtains two elements of the resistivity tensor in the presence of the AHE
\begin{align}
    \rho_{xx} & = \frac{\rho_0}{(1+\tan\theta\tan\vartheta)^2+\tan^2\vartheta} \label{eq.rhoxx} , \\
    \rho_{yx} & = \frac{\rho_0(\tan\vartheta\tan^2\theta+\tan\theta+\tan\vartheta)}{(1+\tan\theta\tan\vartheta)^2+\tan^2\vartheta} . \label{eq.rhoxy}
\end{align}
When both $\tan\theta$ and $\tan\vartheta$ are small, \req{eq.rhoxy} is reduced to \req{eq.empirical} after keeping terms up to the first order. However, if any one of two Hall angles is not negligible, both $\rho_{xx}$ and $\rho_{yx}$ can be substantially different from the Drude resistivity and \req{eq.empirical}, respectively.

\begin{figure}[htbp]
	\begin{center}
		\includegraphics[width=0.9\columnwidth]{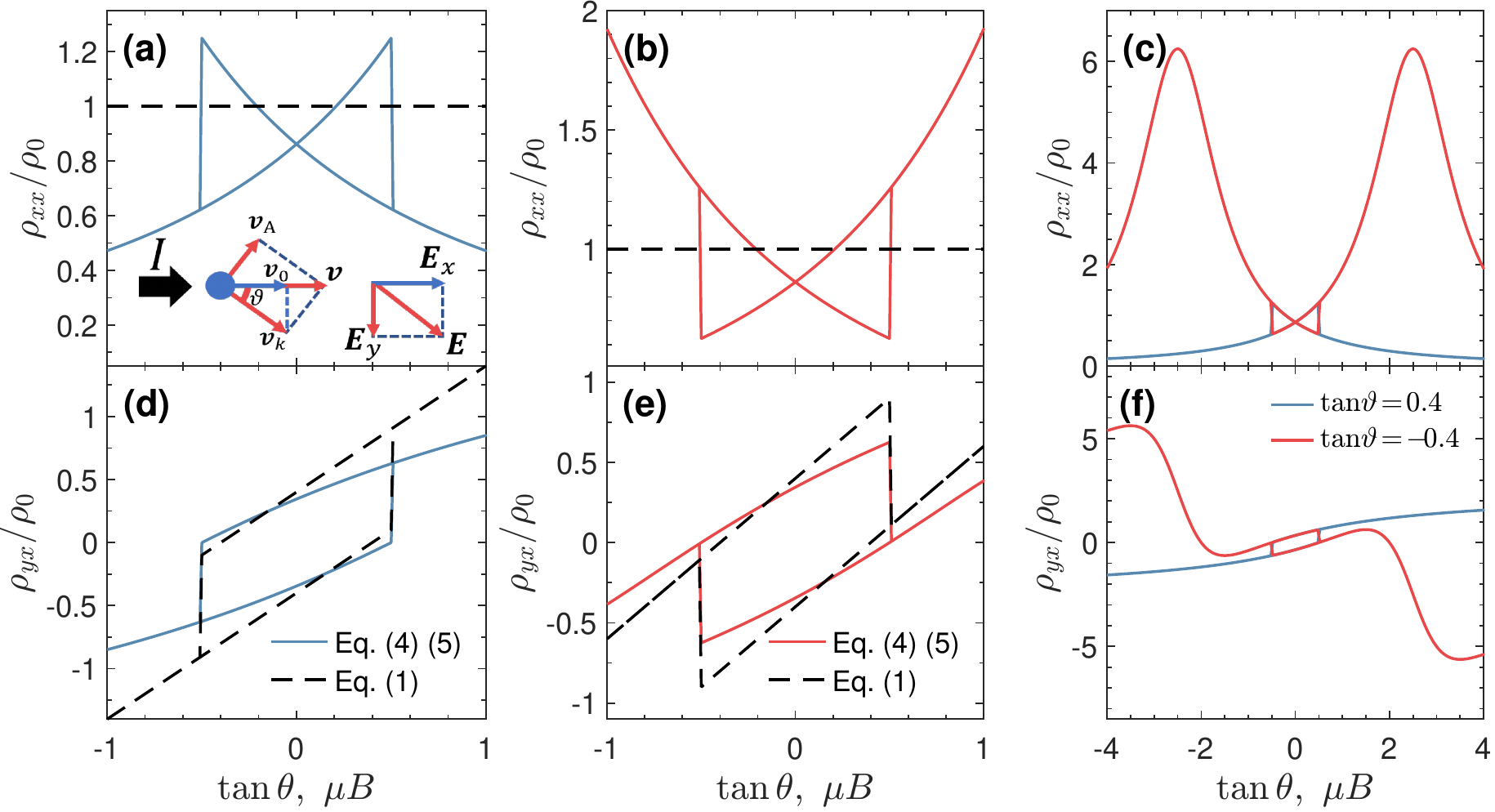}
		\caption{Field dependence of MR and Hall calculated from \reqs{eq.rhoxx} , (\ref{eq.rhoxy}) , and \req{eq.empirical}. The coercive field is set at $\mu B=0.5$. (a),(d) $\tan\vartheta=0.4$ when $\tan\theta>0$. The black dashed lines indicate the Drude resistivity $\rho_0$, which is independent of $B$. (b),(e) $\tan\vartheta=-0.4$ when $\tan\theta>0$.. The black dashed lines are calculated from \req{eq.empirical}. The inset in (a) depicts the carrier velocity $\boldsymbol{v}$ and the electric field $\boldsymbol{E}$. Here, $\boldsymbol{v}_0$ and $\boldsymbol{v_k}$ are the band velocity with and without AHE, respectively. (c),(f) $\rho_{xx}$ and $\rho_{yx}$ in a large $\tan\theta$ (field) range.} 
		\label{fig.simul}
	\end{center}
\end{figure}

In \rfig{fig.simul}, we plot the field dependence of $\rho_{xx}$ and $\rho_{yx}$ calculated from \reqs{eq.rhoxx} and (\ref{eq.rhoxy}). For comparison, the results described by \req{eq.empirical} are also presented. Several unexpected features stand out. Firstly, the zero-field longitudinal resistivity deviates from the Drude resistivity. Secondly, $\rho_{xx}$ is now field dependent, in sharp contrast to the field independent $\rho_0$. Moreover, the field dependence exhibits a strong $B$-linear component ($B$ antisymmetric). The linearity switches sign with $\tan\vartheta$, hence magnetization. As a result, $\rho_{xx}$ exhibits a discontinuity at the coercive field $B_c$. The bowtie feature in MR is similar to that contributed by electron-magnon scattering\cite{Nguyen2011Sep}. However, the magnon MR is usually negative. That is, $\rho_{xx}$ drops when magnetization becomes parallel to field. Here, the sign of MR depends on the sign of $\tan\vartheta$. When $\theta$ and $\vartheta$ have the same signs, the MR is negative. Otherwise, it is positive. As for $\rho_{yx}$, it is not linear in $B$ despite the fact that there is only one isotropic band. The most prominent feature is that when the magnetization remains unchanged, $\rho_{yx}(B)-\rho_{yx}(0)$ is not strictly $B$-antisymmetric, as opposed to \req{eq.empirical}. The deviation from the $B$ antisymmetry is easily recognized from the concave curvature below the coercive field. To see how these terms appear, we show the Taylor series expansion of \reqs{eq.rhoxx} and (\ref{eq.rhoxy}) up to $\tan^2\theta$ 
\begin{align}
\rho_{xx} & = \rho_0 \left( \frac{1}{1+\tan^2\vartheta} - \frac{2\tan\vartheta}{(1+\tan^2\vartheta)^2}\tan\theta+\frac{\tan\vartheta ^2(3-\tan\vartheta ^2)}{{(\tan\vartheta ^2+1)}^3}\tan^2\theta  \right) , \label{eq.rhoxxTaylor}\\
\rho_{yx} & = \rho_0 \left(\frac{\tan\vartheta}{1+\tan^2\vartheta} + \frac{1-\tan^2\vartheta}{(1+\tan^2\vartheta)^2}\tan\theta - \frac{\tan\vartheta(1-3\tan^2\vartheta)}{(1+\tan^2\vartheta)^3}\tan^2\theta  \right) . \label{eq.rhoxyTaylor}
\end{align}
It can be seen that both the $B$-linear term of $\rho_{xx}$ and the $B^2$ term of $\rho_{yx}$ increase with the carrier mobility. When $\tan\vartheta$ is small, these terms are proportional to $\tan\vartheta$. Another noteworthy observation is that all these unexpected components are an odd function of $\tan\vartheta$, which is actually consistent with the Onsager reciprocity relations \cite{Jiang2021Jun}.

Note that there is no MR in an isotropic single band system \cite{Pippard1989}, as indicated by \req{eq.rho}. This is because the Lorentz force is balanced by the force exerted by the Hall electric field in a steady current state. In contrast, AHE generates an anomalous velocity $\boldsymbol{v}_\mathrm{A}$, rather than a force. This velocity, equal to $\frac{e}{\hbar}\boldsymbol{E}\times\boldsymbol{\Omega_k}$ at zero field, is perpendicular to the total electric field $\boldsymbol{E}$ \cite{Xiao2010}, which is tilted away from the direction of the electric current by $\vartheta$. Here, $\boldsymbol{\Omega_k}$ is the Berry curvature. As sketched in the inset of \rfig{fig.simul}(a), the vector sum of this anomalous velocity $\boldsymbol{v}_\mathrm{A}$ and the band velocity $\boldsymbol{v_k}$ is larger than the velocity $\boldsymbol{v}_0$ in the absence of an AHE, leading to a reduction of the longitudinal resistivity. When a magnetic field is applied, the electric field will be gradually rotated, leading to a negative or positive MR depending on the sign of $\vartheta$. When $\tan\theta$ is large and has a sign opposite to $\tan\vartheta$, \reqs{eq.rhoxx} and (\ref{eq.rhoxy}) predict a resistance peak and a sign reversal of the Hall resistance, which will be discussed later.

\begin{figure}[htbp]
	\begin{center}
		\includegraphics[width=0.9\columnwidth]{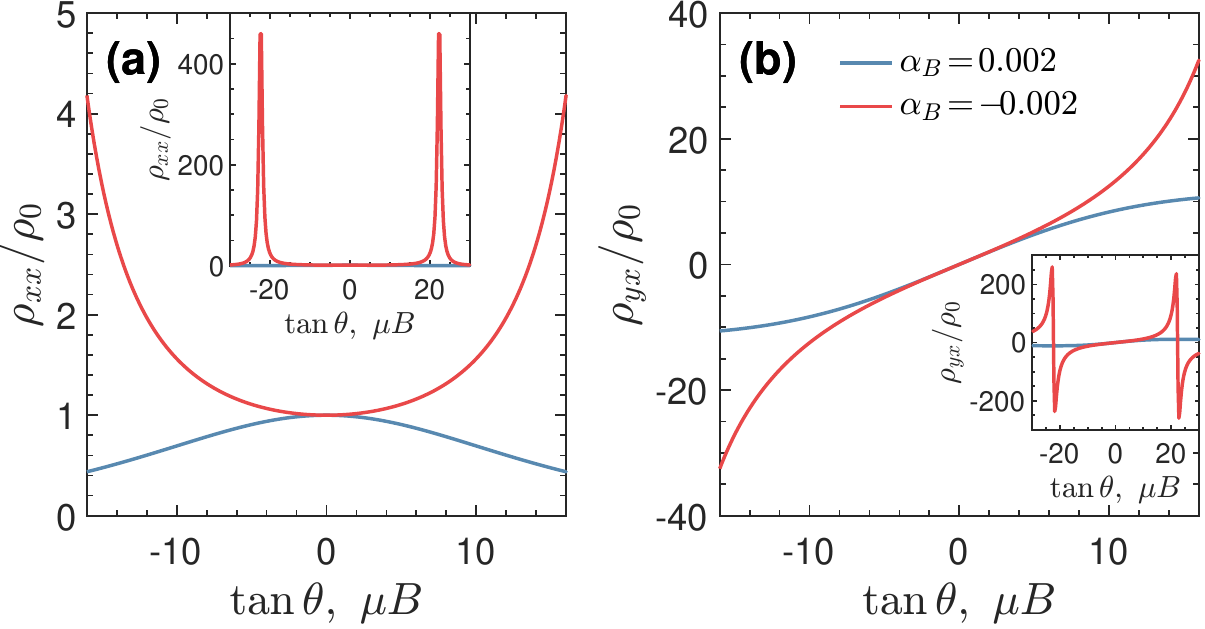}
		\caption{Field dependence of MR and Hall calculated from \reqs{eq.rhoxx} and (\ref{eq.rhoxy}) with $\tan\vartheta$ linear in $B$. $\alpha_B=0.002$ is used, such that when a Zeeman field of 150 T is applied, the anomalous Hall angle is 0.3, close to that in Co$_3$Sn$_2$S$_2$. (a) MR. Data in a large $\tan\theta$ range are shown in the inset. A strong peak appears at $\tan\theta \tan\vartheta\approx -1$. (b) Hall. Data in a large $\tan\theta$ range are shown in the inset. An abrupt sign reversal appears at $\tan\theta \tan\vartheta\approx -1$.}
		\label{fig.extremlyHighField}
	\end{center}
\end{figure}

We now extend the above analysis further to a case in which $\tan\vartheta$ is also linear in $B$. A $B$-linear anomalous Hall effect can appear in paramagnetic\cite{Fert1976,Culcer2003,Zhou2022Sep} and antiferromagnetic materials \cite{Nakatsuji2015,Suzuki2016Dec,Li2022Jul}, but the AHE-induced transport is probably more pronounced in three-dimensional (3D) topological Dirac semimetals, because they often have extremely high carrier mobility. In these materials, each Dirac point may be split into two Weyl cones by a magnetic field along certain crystal directions\cite{Wang2012}. The separation between two Weyl cones in momentum space is proportional to $B$. Since the anomalous Hall conductivity in Weyl semimetals is proportional to the separation\cite{Burkov2014}, one would expect that $\tan\vartheta=\sigma_{xy}^\mathrm{A}/\sigma_0=\alpha_B B$, where $\alpha_B$ is a coefficient. $\alpha_B$ is likely very small, as the applied field in most cases is much smaller than the exchange field that is responsible for AHE in magnetic materials. Owing to their extremely high mobility, a Hall angle much greater than 1 is experimentally accessible. The AHE-induced transport can be significant even when the anomalous Hall angle is very small. Figures \ref{fig.extremlyHighField}(a) and (b) display the simulation results. The MR is quadratic in field, instead of linear at low fields for a constant $\sigma_{xy}^\mathrm{A}$. The Hall resistance is nonlinear, mimicking a two-band behavior. When $\tan\theta$ and $\tan\vartheta$ share the same sign, the field dependence is sublinear, otherwise it is superlinear.

\begin{figure}[htbp]
	\begin{center}
		\includegraphics[width=0.9\columnwidth]{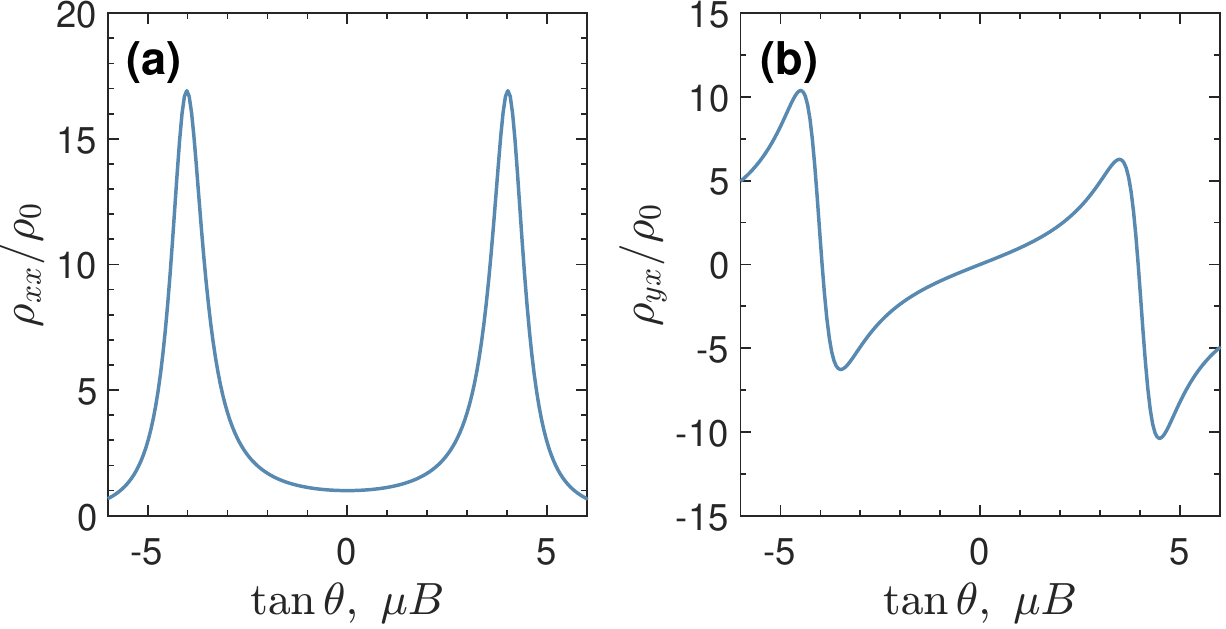}
		\caption{Field dependence of MR and Hall calculated from \reqs{eq.rhoxx} and (\ref{eq.rhoxy}) with $\alpha_B=-0.06$. (a) MR. (b) Hall.}
		\label{fig.similarToZrTe5}
	\end{center}
\end{figure}

When $\tan\vartheta$ is small while $\tan\theta$ is large (strong field) such that $\tan\theta\tan\vartheta\approx -1$, $\rho_{xx}$ may become gigantic, on the order of $\rho_0/\tan^2\vartheta$. $\tan\theta$ can reach a large value in many topological Dirac semimetals of high mobility. For instance, ZrTe$_5$ can have a mobility over $10^5\ \mathrm{cm^2V^{-1}s^{-1}}$ and reach $\tan\theta=1$ at 0.067 T \cite{Tang2019May}. Cd$_3$As$_2$ has a mobility of $8\times 10^6\ \mathrm{cm^2V^{-1}s^{-1}}$ \cite{Liang2015Mar}. With further increase of $\tan\theta$, $\rho_{xx}$ is strongly suppressed. Interestingly, a negative MR following a large positive MR with increasing field has in fact been observed in ZrTe$_5$ \cite{Zhou2019Sep}. The MR simulated from \reqs{eq.rhoxx} and (\ref{eq.rhoxy}) with $\alpha_B=-0.06$ is depicted in \rfig{fig.similarToZrTe5}, which qualitatively reproduces the MR peak observed in experiments. In the meantime, $\rho_{yx}$ can abruptly change sign, which is in contrast to a smooth change in a two-band system. Interestingly, an abrupt change in Hall has been observed in TaP and attributed to Weyl node annihilation \cite{Zhang2017Oct}, but there was no concomitant MR peak. The bipolar behavior can be traced back to the condition that $\theta$ and $\vartheta$ have opposite signs. The overall behavior is similar to that shown in \rfigs{fig.simul}(c) and (f), although the MR can be much stronger because of the smallness of $\alpha_B$. Except for the nonlinear Hall and the quadratic MR at low fields, many features are distinctive from the two-band transport, such as the negative MR, MR peak and sharp reversal of the Hall resistance. The resistivity that asymptotically approaches zero at high fields is particularly surprising. At the same time, $\sigma_{xy}$ asymptotically approaches a plateau of $\sigma_{xy}^\mathrm{A}$ for a constant $\tan\vartheta$. Although this may look like a quantized Hall effect, the plateau value is not universal and bears no relation to any topological invariant. At last, one should keep in mind that the role of Landau levels in high fields is neglected in \req{eq.sigma}. It is possible that these high-field features may be altered. 

Co$_3$Sn$_2$S$_2$ displays a record high anomalous Hall angle and relatively high mobility compared to typical magnetic materials \cite{Liu2018,Yang2020b,Tanaka2020Oct}. A high mobility indicates a strong increase of the Hall angle with magnetic field. In addition, the coercive field in thin films was found to be greatly enhanced \cite{Yang2020b,Tanaka2020Oct,Zeng2021Jul}. These properties make Co$_3$Sn$_2$S$_2$ an ideal place to study the AHE-induced transport described by \reqs{eq.rhoxx} and (\ref{eq.rhoxy}). Indeed, unusual behaviors in both the MR and Hall resistivity have been reported \cite{Yang2020b,Tanaka2020Oct,Zeng2021Jul}.

At low temperatures, the longitudinal resistivity jumps up when the direction of magnetization switches from $B$ antiparallel to parallel at $B_c$\cite{Yang2020b,Tanaka2020Oct,Zeng2021Jul}. The resistivity difference between antiparallel and parallel directions linearly depends on field \cite{Zeng2021Jul}. With increasing temperature, the jump turns into a sudden drop. Yang {\it et al.} argued that the jump at low temperatures was a Lorentz-like ordinary MR due to a fictitious magnetic field induced by magnetization \cite{Yang2020b}, while it was proposed by Zeng {\it et al.} that possible nonlinear magnetic textures and the chiral magnetic field associated with Weyl fermions accounted for the phenomenon \cite{Zeng2021Jul}. We find that \req{eq.rhoxx} naturally explains the above observations, as depicted in \rfig{fig.simul}. The main contribution comes from the $\tan\theta$-linear term in \req{eq.rhoxxTaylor}. Apparently, it is antisymmetric in magnetization ($\tan\vartheta$) and linearly depends on $B$. Note that electron-magnon scattering gives rise to a resistance drop at $B_c$ when the magnetization goes from antiparallel to parallel. The effect increases with the magnon population, hence the temperature. Consequently, the experimentally observed change from a jump to a drop with increasing temperature is expected.

Since the AHE-induced MR is not strong, the bowtie feature depicted in \rfigs{fig.simul}(a) and (b) may not be obvious here. In quantum anomalous Hall insulators, the anomalous Hall angle diverges as the system enters into the quantum Hall state. One would expect a strong manifestation of \req{eq.rhoxx}. Indeed, we find that the bowtie feature appeared in various quantum anomalous Hall insulators \cite{Chang2015May,Deng2020Feb,Ovchinnikov2021Mar,Mogi2022Apr}. The feature may be overwhelmed by the resistivity peak at quantum Hall plateau transitions that coincides with the reversal of the magnetization. Fortunately, MnBi$_2$Te$_4$ thin layers exhibit multiple magnetic transitions with increasing field. The anomalous Hall resistivity displays a significant hysteresis loop only in the first transition at low fields, while the resistivity peak occurs at the highest coercive field. Consequently, a bow-tie-shaped MR was apparent \cite{Deng2020Feb,Ovchinnikov2021Mar}. Furthermore, when the carrier density was tuned by a gate, the relative sign between $\tan\theta$ and $\tan\vartheta$ can be changed. As a result, the resistivity jump at the first coercive field became a drop, while the MR changed from positive to negative (See Fig. S5B in the Supplementary Materials of \rref{Deng2020Feb}). The same correlation between the sign of the jump and the sign of MR was also observed in a semi-magnetic topological insulator (See Fig. S7AB in the Supplementary Materials of \rref{Mogi2022Apr}). All these observations are in excellent agreement with \req{eq.rhoxx}.

We now go back to the Hall resistivity of Co$_3$Sn$_2$S$_2$ nanoflakes. A remarkable feature is that $\rho_{yx}(B)-\rho_{yx}(0)$ below $B_c$ is not strictly antisymmetric in field \cite{Yang2020b,Tanaka2020Oct,Zeng2021Jul}. It was found that the deviation of $\rho_{yx}$ from the expected one is proportional to $B^{1.8}$ at low temperatures\cite{Zeng2021Jul}. The sign of the deviation goes with the direction of magnetization. It was speculated that a change in the magnetic texture might lead to a change of the Weyl point separation\cite{Yang2020b} or a gauge field\cite{Zeng2021Jul}, which was responsible for the observations. We point out that \req{eq.rhoxy} includes $B$-symmetric contributions. Moreover, the leading term is proportional to $B^2$, as shown in \req{eq.rhoxyTaylor}. It is antisymmetric in magnetization ($\tan\vartheta$). All features, including the power index, are consistent with \req{eq.rhoxy}.

\begin{figure}[htbp]
	\begin{center}
		\includegraphics[width=0.9\columnwidth]{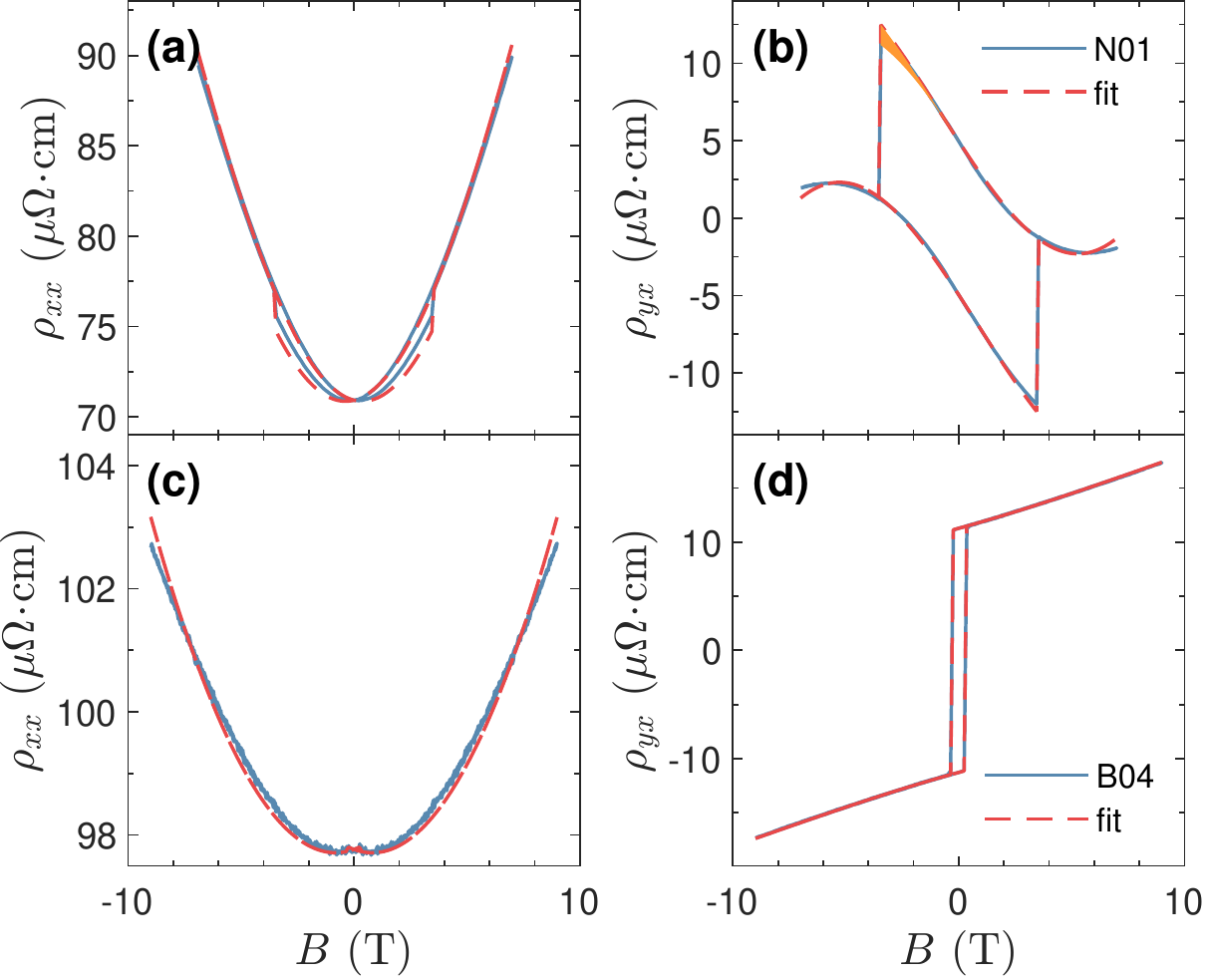}
		\caption{Experimental data of MR and Hall for Co$_3$Sn$_2$S$_2$. (a),(b) $\rho_{xx}$ and $\rho_{yx}$ of N01 at 10~K. A jump in $\rho_{xx}$ occurs at the coercive field. The orange area at negative fields is enclosed by $\rho_{yx}(B)$ and the one obtained by an inversion of $\rho_{yx}(B)$ through $\left(0, \rho_{yx}(0)\right)$, highlighting the $B$-symmetric component of $\rho_{yx}$. (c),(d) $\rho_{xx}$ and $\rho_{yx}$ of B04 at 20~K. Tiny drops in $\rho_{yx}$ at the coercive field are discernible, in agreement with $\tan\theta$ and $\tan\vartheta$ being of the same sign.}
		\label{fig.differentSample}
	\end{center}
\end{figure}

Having shown that previous experimental observations can be qualitatively explained by the AHE, we further carried out transport measurements on Co$_3$Sn$_2$S$_2$ and performed quantitative analysis. Experiment results are shown in \rfig{fig.differentSample}. Benefiting from the large coercive field and high carrier mobility, the nanoflake sample (N01) displays clear AHE-induced MR similar to those reported before. The characteristic $B$-linear term in $\rho_{xx}$ and a $B$-symmetric term in $\rho_{xx}$ are evident. To fit \reqs{eq.rhoxx} and (\ref{eq.rhoxy}) to our data, a two-band model has to be used, as the MR and the nonlinearity in Hall are too strong to be accounted for by a single-band model with AHE. After substituting the one-band Drude conductivity with a two-band one, a reasonably good agreement is obtained. All key features are captured by our model. For comparison, the AHE-induced MR in the bulk crystal is small. Still, the drop in $\rho_{xx}$ can be seen, shown in \rfig{fig.differentSample}(c). This is due to the small coercivity and lower carrier mobility of bulk crystals, which is consistent with \reqs{eq.rhoxxTaylor} and (\ref{eq.rhoxyTaylor}).

\section{Conclusion}

By pointing out the limitation of the empirical resistivity relation that is widely used for AHE data analysis, we emphasize the necessity of employing the proper conductivity relation when either the ordinary Hall angle or the anomalous Hall angle is not small. It is shown that AHE gives rise to a $B$-linear MR in the longitudinal resistivity. The sign of the linearity depends on the sign of the anomalous Hall angle, yielding a characteristic bow-tie-shaped MR. Additionally, AHE induces a nonlinearity and a $B^2$ component in the field dependence of Hall resistivity. The AHE-induced transport can not only explain some MR features in quantum anomalous Hall insulators but also quantitatively account for the unusual observations made in Co$_3$Sn$_2$S$_2$. We further extend the analysis to nonmagnetic 3D topological Dirac semimetals and discuss the possible origin of the large MR observed in these materials based on AHE.

\begin{acknowledgements}
This work was supported by National Key Basic Research Program of China (Grants No. 2022YFA1403700, No. 2020YFA0308800) and NSFC (Projects No. 11774009, No. 12074009, and No. 12204520).  Zhilin Li is grateful for the support from the Youth Innovation Promotion Association of the Chinese Academy of Sciences (No. 2021008).
\end{acknowledgements}

\clearpage

\end{document}